\documentclass[showpacs,aps,prb,twocolumn,superscriptaddress,footinbib]{revtex4} 

\usepackage{graphicx,amssymb,amsmath} 
\usepackage{dcolumn}
\usepackage{bm}

\begin{document}

\title{Existence of an upper limit on the density of excitons in carbon nanotubes by diffusion-limited exciton-exciton annihilation: Experiment and theory}

\author{Yoichi Murakami}
\email[]{ymurak@chemsys.t.u-tokyo.ac.jp}
\thanks{corresponding author.}
\affiliation{Department of Electrical and Computer Engineering, Rice University, Houston, Texas 77005, USA}
\affiliation{Department of Chemical System Engineering, University of Tokyo, 
Bunkyo-ku, Tokyo 113-8656, Japan}

\author{Junichiro Kono}
\affiliation{Department of Electrical and Computer Engineering, Rice University, Houston, Texas 77005, USA}

\date{\today}

\begin{abstract}
Through an investigation of photoemission properties of highly-photoexcited single-walled carbon nanotubes, we demonstrate that there is an upper limit on the achievable excitonic density.  As the intensity of optical excitation increases, all photoluminescence emission peaks arising from different chirality single-walled carbon nanotubes showed clear saturation in intensity.  Each peak exhibited a saturation value that was independent of the excitation wavelength, indicating that there is an upper limit on the excitonic density for each nanotube species.  We propose that this saturation behavior is a result of efficient exciton-exciton annihilation through which excitons decay non-radiatively.  In order to explain the experimental results and obtain excitonic densities in the saturation regime, we have developed a model, taking into account the generation, diffusion-limited exciton-exciton annihilation, and spontaneous decays of one-dimensional excitons.  Using the model, we were able to reproduce the experimentally obtained saturation curves under certain approximations, from which the excitonic densities were estimated.  The validity of the model was confirmed through comparison with Monte Carlo simulations.  Finally, we show that the conventional rate equation for exciton-exciton annihilation without taking into account exciton diffusion fails to fit the experimentally observed saturation behavior, especially at high excitonic densities.

\end{abstract}


\maketitle

\section{Introduction}

The optical and electronic properties of low-dimensional materials have been an important subject of study in the field of condensed matter physics.  In particular, one-dimensional (1-D) materials are predicted to possess unique properties that are distinctly different from those at higher dimensions,\cite{Ogawa-book,Giamarchi-book} primarily due to the enhanced Coulomb interactions among the quantum confined charge carriers.  One common feature of optically-excited low-dimensional systems is the formation of strongly bound electron-hole ($e$-$h$) pairs, or excitons,\cite{Knox-book} which dominate interband optical spectra.  1-D semiconductors are expected to show an almost complete suppression of optical absorption at the band edges, with a significant fraction of the total oscillator strength taken by the lowest excitonic state.\cite{Ogawa-PRB-1991-both,Rossi-PRL-1996}

Early reports of lasing from semiconductor quantum wires (QWRs)\cite{Kapon-PRL-1989,Wegscheider-PRL-1993} invoked much interest in the physics of \emph{high density} 1-D excitons.  A number of studies have thus far been performed on such QWR systems during the last two decades to understand many-body phenomena (e.g., lasing, band-gap renormalization, biexciton formation, and the Mott transition),\cite{Ambigapathy-PRL-1997,DasSarma-PRL-2000,DasSarma-PRB-2001,Rubio-SSC-2001,Akiyama-PRB-2003,Guillet-PRB-2003,Yoshita-PRB-2006,Hayamizu-PRL-2007} but many aspects are still under debate and not well understood.  More recently, studies on high-density $e$-$h$ pairs have been extended to novel 1-D materials such as conjugated polymers\cite{Xu-PRB-2003,Dicker-PRB-2005,Zaushitsyn-PRB-2007,King-PRB-2007} and single-walled carbon nanotubes (SWNTs).\cite{Wang-PRB-2004,Ostojic-PRL-2005,Ma-PRL-2005}  The latter are tubular materials made of $sp^2$-bonded carbon atoms,\cite{Iijima-Nature-1993} attracting much recent interest from diverse research fields due to their unique properties.\cite{Dresselhaus-book}  Semiconducting SWNTs are known to have extremely strong quantum confinement of $\sim$ 1~nm, giving rise to large exciton binding energies on the order of 0.5-1~eV,\cite{Spataru-PRL-2004,Wang-Science-2005,Maultzsch-PRB-2005} much larger than those of GaAs QWRs ($\sim$~20~meV)\cite{Rossi-PRL-1996,Wegscheider-PRL-1993} and comparable to or larger than those of conjugated polymers ($\sim$~0.4~eV\cite{King-PRB-2007} and $<$~0.1~eV\cite{Moses-PNAS-2001}).

Here, we report results of experimental and theoretical investigations on the properties of photoluminescence (PL) from excitons in SWNTs through nonlinear photoluminescence excitation (PLE) spectroscopy using intense optical pulses.  From the clear saturation behavior observed in the intensities of all the PL features as a function of excitation laser intensity as well as the complete {\em flattening} of the PLE spectra observed at very high laser intensities, we show the existence of an upper limit on the density of excitons that can be accommodated in SWNTs. Such an upper limit is considered to be caused by the diffusive motion of the excitons\cite{Sheng-PRB-2005,Russo-PRB-2006,Cognet-Science-2007} combined with highly rapid and efficient exciton-exciton annihilation (EEA) in SWNTs.\cite{Ma-PRL-2005}
As described in Section \ref{Model}, we have developed a theoretical model for describing diffusion-limited EEA processes
in 1-D, which enabled us to simulate the PL saturation curves and estimate the densities of excitons in SWNTs as a function of excitation intensity.

A portion of this work was described in our earlier letter.~\cite{murakami-under-review}  The purpose of the present paper is to provide a complete description of both experimental and theoretical aspects of this study.

\section{Experiment}
\label{experiment}

\subsection{Experimental methods}

The sample was prepared by ultrasonicating CoMoCAT SWNTs in D${_2}$O with 1 wt\% sodium cholate for 1 hour, followed by ultracentrifugation at 111,000 g for 4 hours. This centrifugation condition is sufficient to remove SWNT bundles effectively.\cite{Tan-JPCB-2005} Only the upper 50 \% of the supernatant was collected and used for the experiment. 
The solution was put in a 1-mm-thick quartz cuvette.  The optical density of the sample around the $E_{22}$ resonance was below 0.2, which helped avoid non-uniform excitation and re-absorption of the emitted PL within the sample.  The excitation source was an optical parametric amplifier (OPA), producing $\sim$~250~fs pulses at a repetition rate of 1~kHz, tunable in the visible and near-infrared ranges, pumped by a chirped-pulse amplifier (Clark-MXR, Inc., CPA-2010).  Optical filters were carefully selected and set in the beam path to thoroughly eliminate any parasitic wavelength components (mostly in the ultraviolet and near-infrared regions) contained in the OPA beam.  The OPA beam was focused onto the sample to a spot size of 300-400~$\mu$m.  Only the central $\sim$~2~mm portion of the OPA beam profile ($\sim$~6~mm) was taken out by using an aperture just before the focus to enhance the spot uniformity at the sample.  The PL from the sample was focused onto the monochromator entrance and recorded with a liquid-nitrogen-cooled InGaAs 1-D array detector.  The obtained PL spectra were corrected for the wavelength-dependence of the grating efficiency and detector sensitivity.

For the data shown in Fig.~\ref{PL_change}(c), a different sample as well as a different excitation light source was used for verifying the universality of the phenomena observed.  In this case, the sample was a dried film of CoMoCAT SWNTs embedded in $\iota$-carrageenan, formed by drying a mixed gel of $\iota$-carrageenan and the centrifuged supernatant of CoMoCAT SWNTs on the surface of a sapphire substrate.  The sapphire substrate served as a mechanical support as well as a heat sink of the film during the measurements.  The excitation light was 1~kHz and $\sim$~250~fs optical pulses with a central wavelength of 653~nm (FWHM = 10~nm), produced by filtering whitelight pulses generated by focusing the CPA beam onto a sapphire crystal.

\subsection{Experimental results}
\label{exp-results}

\begin{figure} [tbp]
\includegraphics[scale=0.70]{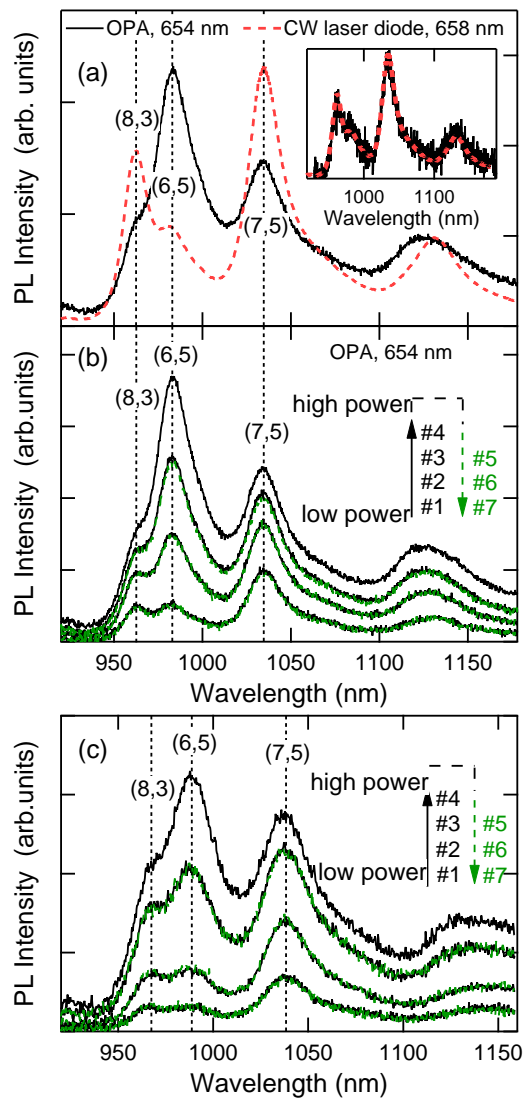}
\caption{(color online) Pump-intensity-dependent PL spectra measured for (a, b) the centrifuged supernatant of CoMoCAT SWNTs dispersed in D$_2$O and (c) CoMoCAT SWNTs embedded in a dried $\iota$-carrageenan film: (a)~Black solid --- spectrum obtained with OPA pulse (654~nm, 29~nJ).  Red dotted curve --- spectrum obtained with a CW laser diode (658~nm, 100~$\mu$W).  Inset shows that the two spectra coincide when the OPA pulse energy is very low (300~pJ).  (b)~Change of PL spectra with the pulse energy of OPA beam (654~nm) varied between 1~nJ and 30~nJ in the order of \#1 to \#7.  Curve \#4 corresponds to the highest fluence ($\sim$1.3~$\times$~10$^{14}$~photons/cm$^2$).  (c)~Change of PL spectra with the pulse energy of 653~nm light (FWHM = 10~nm) varied between 1~nJ and 20~nJ in the order of \#1 to \#7.}
\label{PL_change}
\end{figure}

Figure~\ref{PL_change}(a) compares two PL spectra.  The black solid curve was obtained using the OPA with a wavelength of 654~nm (or 1.90~eV) and a pulse energy of 29~nJ, while the red dotted curve was obtained using a weak (100~$\mu$W) CW laser with a wavelength of 658~nm (or 1.88~eV).  It is seen that the relative intensities of different PL peaks are drastically different between the two curves.  The inset confirms that the two spectra coincide accurately when the OPA pulse energy was kept very low (300~pJ).  Figure~\ref{PL_change}(b) shows PL spectra measured with pulse energies of 1~nJ (curves \#1 and \#7), 4~nJ (\#2 and \#6), 10~nJ (\#3 and \#5), and 30~nJ (\#4).  The (7,5) peak is dominant at low fluences while the (6,5) peak becomes dominant at high fluences.  It is important to note that the different curves were taken in the order of \#1 to \#7, demonstrating that the observed changes are reproducible and are not caused by any laser-induced permanent change in the sample.  Additionally, note that the PL intensities tend to saturate at high laser fluences, while their peak positions do not change at all.

Figure~\ref{PL_change}(c) shows PL spectra measured for the dried $\iota$-carrageenan film using 653~nm optical pulses with a FWHM bandwidth of 10~nm at different pulse energies.  The pulse energies were 1~nJ (curves \#1 and \#7), 3~nJ (\#2 and \#6), 10~nJ (\#3 and \#5), and 20~nJ (\#4) measured in the order of \#1 to \#7.  Figure~\ref{PL_change}(c) exhibits the same behavior as that shown in Fig.~\ref{PL_change}(b), demonstrating that the observed changes shown in Fig.~\ref{PL_change}(b) did not result from any artifacts, e.g., caused by the fluidic nature of the sample or by the unnoticed parasitic wavelength components in the OPA beam.  In the following, we use the excitation pulse fluence in terms of the number of incident photons per cm$^2$ per pulse to express the intensity of excitation pulses.

\begin{figure} [tbp]
\includegraphics[scale=0.6]{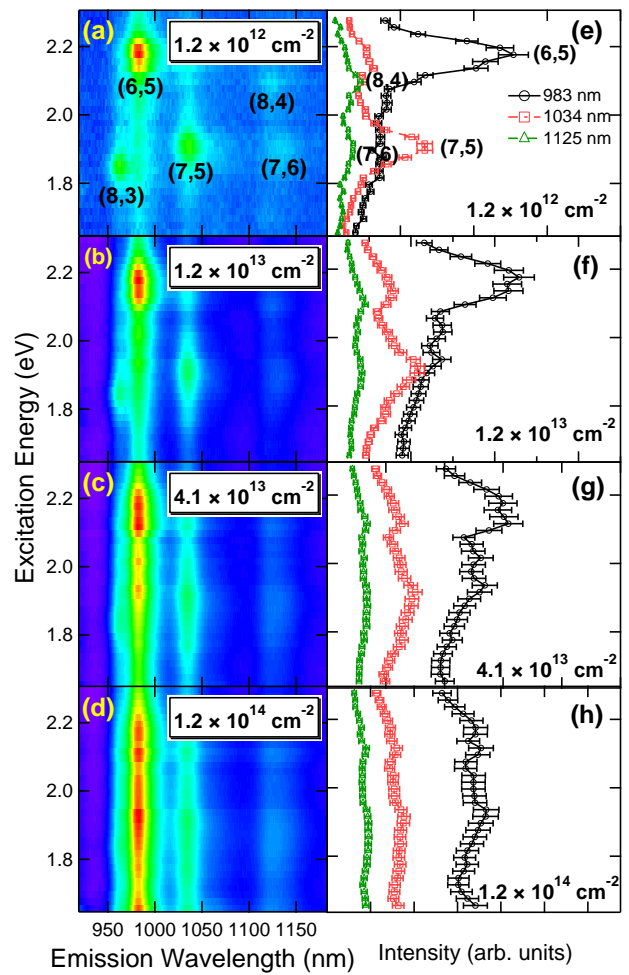}
\caption{(color online) Evolution of PLE data with increasing pump pulse fluence: (a) 1.2 $\times$ 10$^{12}$, (b) 1.2 $\times$ 10$^{13}$, (c) 4.1 $\times$ 10$^{13}$, and (d) 1.2 $\times$ 10$^{14}$ photons/cm$^2$.
(e-h): PLE spectra corresponding to (a)-(d) at emission wavelengths of 983~nm (circles), 1034~nm (squares), and 1125~nm (triangles).}
\label{NL-PLE}
\end{figure}

Figures \ref{NL-PLE}(a)-\ref{NL-PLE}(d) show PLE maps taken with various pump fluences.  The step size for the pump photon energy was 20~meV.  The data taken with the lowest fluence (1.2~$\times$~10$^{12}$ photons/cm$^2$) [Fig.~\ref{NL-PLE}(a)] is essentially the same as that taken with low-intensity CW light.  However, as the fluence is increased [Figs.~\ref{NL-PLE}(b)-\ref{NL-PLE}(d)], the $E_{22}$ excitation peaks gradually broaden and eventually become completely flat at the highest fluence (1.2~$\times$~10$^{14}$~photons/cm$^2$) --- i.e., {\em PL intensities become independent of the excitation wavelength}.  The corresponding PLE spectra are shown in Figs.~\ref{NL-PLE}(e)-\ref{NL-PLE}(h) for three PL wavelengths at 983, 1034, and 1125~nm.  Again, such changes in the PLE spectra were reproducible over the fluence range tested here, indicating that no sample damage was induced.

In order to obtain PL intensity ($I_{\rm PL}$) versus pump intensity ($I_{\rm pump}$) relationships for different emission peaks, we measured PL spectra at different photon fluences for various excitation wavelengths.  Each PL spectrum was decomposed and fitted by multiple peaks corresponding to the SWNT types/chiralities involved in the measured wavelength range.  50\% Gaussian + 50\% Lorentzian lineshape was assumed, and the decomposition was performed by optimizing the peak-width so that the decomposition gives the best fitting for the original PL spectrum.  The optimum widths at the highest fluence was larger by $\sim$~15~\% than those at the lowest fluence in Fig.~\ref{PL_change}, and such an increment of the width is considered to be caused by the enhanced interactions among excitons or their reduced lifetime in the presence of high density excitons.  Throughout the decomposition analysis performed, peak positions of all the PL features and the ratios among their widths were fixed regardless of the excitation wavelength and fluence.

\begin{figure} [tbp]
\includegraphics[scale=0.68]{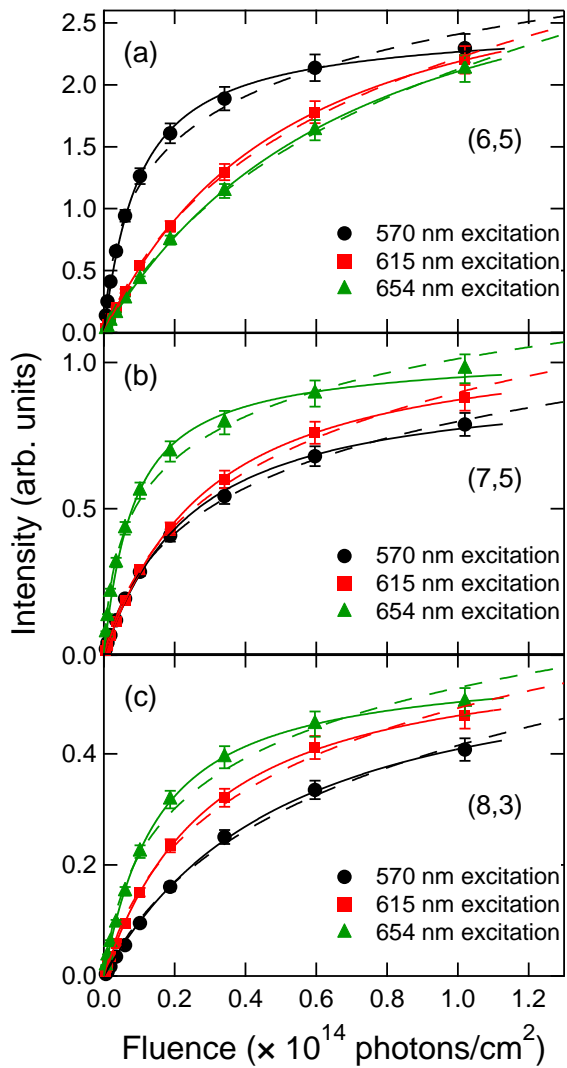}
\caption{(color online)  Integrated PL intensity versus pump fluence for (6,5), (7,5), and (8,3) SWNTs.  Pump wavelengths were 570~nm (circles), 615~nm (squares), and 654~nm (triangles).  The error bars account for $\pm$5\%.
The solid and dashed curves are fitting by Eq.~(\ref{instantaneous-eqn_(lambda=1)}) and Eq.~(\ref{conventional_3}), respectively.}
\label{saturation}
\end{figure}

Figure~\ref{saturation} shows the obtained integrated PL peak intensities ($I_{\rm PL}$) plotted against the incident photon fluence ($I_{\rm pump}$) for (6,5), (7,5), and (8,3) SWNTs at excitation wavelengths of 570, 615, and 658~nm.
The resonance wavelengths of these SWNT types at the $E_{22}$ levels are approximately 570, 647, and 673~nm, respectively.  It can be seen that the integrated PL intensity begins to saturate at a lower (higher) fluence when SWNTs are excited resonantly (non-resonantly). Unexpectedly fast saturation of the PL from (7,5) with 570~nm excitation (which is non-resonant) can be attributed to its proximity to the phonon sideband at $\sim$~585~nm.\cite{Miyauchi-PRB-2006}

\subsection{Data interpretation}

We interpret these observations as results of very efficient exciton-exciton annihilation,\cite{Suna-PRB-1970,Xu-PRB-2003,Dicker-PRB-2005,Ma-PRL-2005,Zaushitsyn-PRB-2007,King-PRB-2007} a non-radiative process that occurs at high exciton densities where two excitons are spatially close enough to interact with each other, resulting in the annihilation of the two excitons and simultaneous creation of an \emph{e-h} pair in a higher energy state (either as a bound exciton or an unbound free \emph{e-h} pair).  We assume that the formation of $E_{11}$ excitons occurs in a very short time scale after an optical excitation around $E_{22}$, because of much faster $E_{22}$-to-$E_{11}$ relaxation (e.g., $\sim$40~fs~\cite{Manzoni-PRL-2005}) than the duration of our OPA pulse ($\sim$250~fs).  Thus, excitons quickly accumulate in the $E_{11}$ state during and right after photo-creation of $e$-$h$ pairs.  However, the number of excitons that can be {\em accommodated} in the $E_{11}$ state is limited by EEA.  As the exciton density, $n_{\rm x}$, approaches its maximum value, EEA begins to prevent a further increase by {\em efficiently removing excitons non-radiatively}, which explains the PL saturation behavior.  Since EEA serves as a bottleneck for the exciton density, the PL intensity becomes insensitive to whether the excitons were created resonantly or non-resonantly around the $E_{22}$ level and independent of the pump wavelength, resulting in the flattening of PLE spectra.  Namely, at very high pump fluences, the PL intensity is determined not by how efficiently excitons are created but by {\em how many $E_{11}$ excitons can be accommodated within a particular type of SWNT} as well as by the relative abundance of that type of SWNT in the sample. We also performed optical transmission spectroscopy in the $E_{22}$ range using OPA pulses and found that the absorption spectra do not exhibit any change even at high pulse fluences.  Thus, nonlinear optical effects such as phase-space filling in the $E_{22}$ range are not playing any role in the observed PLE broadening/flattening and PL saturation.

\section{Theory}
\label{Model}

In this section, we develop a theoretical model for explaining the experimental results, taking into account the generation, diffusion-limited EEA, and spontaneous decays of 1-D excitons. 
Under certain approximations, the model provides a direct analytical relationship between the intensity of the excitation light ($I_{\rm pump}$) and that of the emitted PL ($I_{\rm PL}$) for limiting cases, which allows us to estimate the density of excitons in SWNTs through fitting to the experimentally obtained $I_{\rm pump}$ vs.~$I_{\rm PL}$ curves.

\subsection{Model}

\begin{figure}
\includegraphics[scale=0.49]{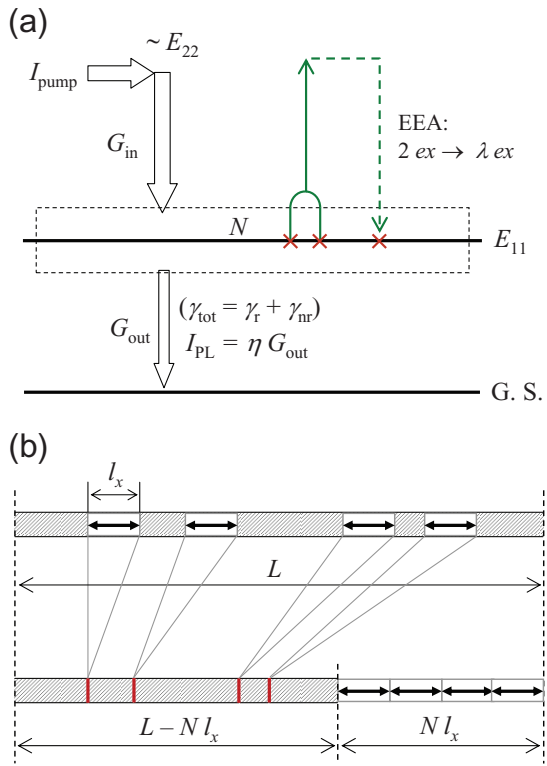}
\caption{(color online) (a) Schematic energy diagram of the system considered in the model.  The dotted box enclosing the lowest energy level ($E_{\rm 11}$) is the domain of interest where $N$ excitons are populated.  All the symbols are defined in the text.  (b) Schematic description of $N = 4$ excitons randomly distributed over a SWNT with a length of $L$.  The horizontal arrow with a length of $l_{\rm x}$ denotes the average length traversed by one exciton during its spontaneous decay lifetime $\tau_{\rm tot}$.  The lower part is the equivalent of the upper but emphasizes that the total length of the unoccupied region is $L - N l_{\rm x}$ where $N$ vertical thick bars denote the borders of the areas occupied by those excitons.  The ends of the SWNT are assumed to be a cyclic boundary.}
\label{schematic_1}
\end{figure}

Figure~\ref{schematic_1}(a) shows a schematic energy diagram of the excitons in semiconducting SWNTs under consideration.  
We are interested in calculating the population of excitons $N$ (indicated by the dotted box) in the lowest energy state $E_{\rm 11}$. First, excitons are created by optical excitation at an energy around the $E_{\rm 22}$ level, which is typically in the visible wavelength range.  The excitation intensity is denoted by $I_{\rm pump}$.  As soon as excitons are created, they decay to the $E_{\rm 11}$ level within a very short time ($\sim$40~fs)\cite{Manzoni-PRL-2005} by transferring their energies to the lattice. Recent studies have reported that the excitons created around $E_{\rm 22}$ levels primarily decay to the $E_{\rm 11}$ level with a probability close to unity.\cite{Hertel-NanoLett-2008,Lebedkin-PRB-2008}  The influx of excitons to the $E_{\rm 11}$ level is denoted by $G_{\rm in}$.  On the other hand, the spontaneous decay time $\tau_{\rm tot}$ of the $E_{\rm 11}$ excitons to the ground state (G.~S.) has been reported to be 10--100~ps.\cite{Ostojic-PRL-2004,Wang-PRL-2004,Reich-PRB-2005}  Such a spontaneous decay consists of radiative and non-radiative processes, with respective rates $\gamma_{\rm r}$ and $\gamma_{\rm nr}$ (s$^{-1}$), where $\gamma_{\rm r} \, + \, \gamma_{\rm nr} \equiv \gamma_{\rm tot} = {\tau_{\rm tot}}^{-1}$.  The outflux of excitons from the $E_{\rm 11}$ level \emph{via the spontaneous decay process} is denoted by $G_{\rm out}$.  Therefore, the flux of the PL photons or the PL intensity ($I_{\rm PL}$) is equal to $\eta G_{\rm out}$, where $\eta$ ($\equiv \gamma_{\rm r}/\gamma_{\rm tot}$) denotes the branching ratio for the radiative decay from the $E_{\rm 11}$ level.

As the density of excitons increases, the EEA process becomes important.  If the \emph{e-h} pair created in the higher energy state returns back to the $E_{\rm 11}$ exciton level with a probability of $\lambda$ ($ 0 \le \lambda \le 1$), the initial two excitons are eventually reduced to $\lambda$ excitons (as an expectation value) through such an EEA process.  We make the following two assumptions: (1) Once two excitons intersect in a SWNT, EEA occurs instantaneously with a probability of one, and (2) the positions where excitons are created in SWNTs by optical excitations are random.

Figure~\ref{schematic_1}(b) schematically shows a situation in which $N$ excitons exist in a SWNT of length of $L$. Each exciton is considered to ``occupy'' a characteristic length $l_{\rm x}$ in a SWNT.  In a static limit, $l_{\rm x}$ should simply be the exciton size. On the other hand, when exciton diffusion is present, $l_{\rm x}$ is assumed to be the average distance traveled by an exciton during the spontaneous decay lifetime $\tau_{\rm tot}$ and is $\propto\sqrt{D \tau_{\rm tot}}$, where $D$ (cm$^{2}$/s) is the exciton diffusion constant.  Namely, it is assumed that $l_{\rm x}$ is determined by the diffusion length, and any two excitons created within $l_{\rm x}$ undergo EEA.

First, a 1-D space of length $L$ is considered where $N$ segments (or excitons) of length $l_{\rm x}$ are randomly present without overlapping each other, as shown in Fig.~\ref{schematic_1}(b).  The probability of a new segment (of length $l_{\rm x}$) to enter the system \emph{without} overlapping any of the $N$ existing segments, $p(N)$, is given by the product of the following two probabilities
\begin{equation}
p_1 = 1 - \frac{N l_{\rm x}}{L}
\label{p1}
\end{equation}
which is the probability for the center of the new segment to land on an unoccupied area [hatched regions in Fig.~\ref{schematic_1}(b)], and
\begin{equation}
p_2 = \left(1 - \frac{l_{\rm x}}{L - N l_{\rm x}} \right)^{N}
\label{p2}
\end{equation}
which is the probability that the occupying length ($l_{\rm x}$) of the new segment (whose center has landed on an unoccupied area) does not interfere with any of the $N$ existing segments [vertical thick bars in the lower part of Fig.~\ref{schematic_1}(b)].  Hence, $p(N)$ is written as
\begin{equation}
p(N) = \left(1 - \frac{N l_{\rm x}}{L}\right) \left(1 - \frac{l_{\rm x}}{L - N l_{\rm x}} \right)^{N} \;.
\label{p_N}
\end{equation}
The expectation value of the increment of $N$ due to the introduction of a new segment into the system, ${\left< \Delta N \right>}_N$ ($0 < {\left< \Delta N \right>}_N \leq 1$), depends on the type of two-particle annihilation considered.  For the general ``$ex + ex \rightarrow \lambda\, ex$'' case, ${\left< \Delta N \right>}_N$ is expressed as
\begin{equation}
{\left< \Delta N \right>}_N = p(N) - (1 - \lambda)(1 - p(N)) \, .
\label{N_N}
\end{equation}
In the following, the derivation for the case of $\lambda = 1$ is shown as a specific example, since this case is considered to represent our experimental situation.  The final results will be presented for both the $\lambda = 1$ and $\lambda = 0$ cases.

\subsection{Solutions in limiting cases}

\subsubsection{Steady-state limit}

Here, we consider the {\em steady-state} limit, where the number of excitons $N$ in the system is steady, corresponding to CW excitation.  In order to derive the relationship between $I_{\rm pump}$ and $I_{\rm PL}$, we consider the relationship between $G_{\rm in}$ and $G_{\rm out}$ [see Fig.~\ref{schematic_1}(a)], where both $G_{\rm in}$ and $G_{\rm out}$ are \emph{rates}, having units of s$^{-1}$.  When $N$ is sufficiently small and EEA is negligible, $G_{\rm in} = G_{\rm out} = \gamma_{\rm tot} N$.  When EEA is present, however, this relationship becomes
\begin{equation}
G_{\rm in} = \frac{G_{\rm out}}{{\left< \Delta N \right>}_N} = \frac{\gamma_{\rm tot}N}{{\left< \Delta N \right>}_N} \, .
\label{G_in}
\end{equation}
In the case of $\lambda = 1$, Eq.~(\ref{N_N}) and Eq.~(\ref{G_in}) lead to
\begin{equation}
G_{\rm in} = \frac{\gamma_{\rm tot} N}
{
\left(
1 - \frac{N l_{\rm x}}{L}
\right)
\left\{
1 - \frac{l_{\rm x}}{L} {\left(1 - \frac{N l_{\rm x}}{L}\right)}^\text{-1}
\right\}^{N}
}  \, .
\label{I_pump_beta}
\end{equation}
We introduce two dimensionless variables $\zeta \equiv N l_{\rm x}/L$ and $\psi \equiv G_{\rm in} l_{\rm x}/\gamma_{\rm tot} L$, the former of which is a dimensionless exciton population ($0 \leq \zeta < 1$).  Using these variables, Eq.~(\ref{I_pump_beta}) can then be rewritten in dimensionless form:
\begin{equation}
\psi = \frac{\zeta}
{
(1 - \zeta)
\left\{
1 - \frac{l_{\rm x}}{L} {\left(1 - \zeta\right)}^{-1}
\right\}^{\frac{L}{l_{\rm x}} \zeta}
}  \, .
\label{psi-zeta1}
\end{equation}
Expanding the second factor in the denominator of Eq.~(\ref{psi-zeta1}) and eliminating higher-order terms of $l_{\rm x} / L$ leads to an equation of only $\zeta$ and $\psi$, expressed as
\begin{equation}
\psi = \frac{\zeta}
{
(1 - \zeta)
\sum\limits_{\kappa=0}^\infty
{
\frac{(-1)^{\kappa}}{\kappa !}
\left(
\frac{\zeta}{1 - \zeta}
\right)^{\kappa}
}
}.
\label{psi-expansion}
\end{equation}
Summing up to $\kappa = 6$ in Eq.~(\ref{psi-expansion}) is sufficient to reproduce Eq.~(\ref{psi-zeta1}) for $l_{\rm x} / L < 0.1$.  Finally, noting that the denominator of Eq.~(\ref{psi-expansion}) is equal to the Taylor expansion of an exponential function, the solution for the steady-state limit for $\lambda = 1$ ($ex + ex \rightarrow ex$) is expressed as
\begin{equation}
\psi = \frac{\zeta}{1-\zeta} \, \mathrm{exp}\left(\frac{\zeta}{1-\zeta} \right) 
\, .
\label{steady-state-eqn_(lambda=1)}
\end{equation}
Here, since $I_{\rm PL} \propto N$ and $I_{\rm pump} \propto G_{\rm in}$, $I_{\rm PL}$ and $I_{\rm pump}$ are proportional to $\zeta$ and $\psi$, respectively, i.e., $I_{\rm PL} = c_1 \zeta$ and $I_{\rm pump} = c_2 \psi$, where $c_1$ and $c_2$ are real constants.

On the other hand, the solution for the case of $\lambda = 0$ ($ex + ex \rightarrow 0$) is derived through a similar procedure, yielding
\begin{equation}
\psi =
\frac{\zeta}
{
2(1 - \zeta) \mathrm{exp}\left( -\frac{\zeta}{1-\zeta} \right) - 1
}  \, .
\label{steady-state-eqn_(lambda=0)}
\end{equation}
Equations~(\ref{steady-state-eqn_(lambda=1)}) and (\ref{steady-state-eqn_(lambda=0)}), valid for CW PL experiments, are implicit equations relating the PL intensity ($I_{\rm PL}$) and the pump intensity ($I_{\rm pump}$) in terms of their respective dimensionless variables $\zeta$ and $\psi$. These equations contain no fitting parameters except the two linear scaling factors $c_1$ and $c_2$ and simply become equivalent ($\zeta = \psi$) in the low density limit ($\zeta \rightarrow 0$).

\subsubsection{Instantaneous limit}

Here, we consider the {\em instantaneous} limit in which creation of all the excitons by an infinitesimally short optical pulse and their internal relaxation to $E_{\rm 11}$ level occur instantaneously at $t = 0$, before diffusion-limited EEA and spontaneous decay processes begin to occur subsequently. This limit represents a situation where the duration of the optical pulse and the time required for intraband relaxation are much shorter than the spontaneous decay time $\tau_{\rm tot}$, as in the case of the present experimental situation. The pump intensity $I_{\rm pump}$ in this limit is directly proportional to $N_0$, the number of excitons or $e$-$h$ pairs created at $t = 0$, while the PL intensity $I_{\rm PL}$ is proportional to the number of excitons $N$ that survived EEA.  The relationship between $N_0$ and $N$ is described by the differential equation
\begin{equation}
d{N_0} = \frac{dN}{{\left< \Delta N \right>}_N}
\label{dN_0}
\end{equation}
where ${\left< \Delta N \right>}_N$ is given by Eq.~(\ref{N_N}).  When $\lambda = 1$ is assumed, Eq.~(\ref{dN_0}) is rewritten as
\begin{equation}
\frac{d N_0}{d N} = \frac{1}
{
\left(
1 - \frac{N l_{\rm x}}{L}
\right)
\left\{
1 - \frac{l_{\rm x}}{L} {\left(1 - \frac{N l_{\rm x}}{L}\right)}^{-1}
\right\}^{N}
} \, .
\label{dN0_dN}
\end{equation}
As in the previous case (steady-state limit), we introduce dimensionless variables $\zeta \equiv N l_{\rm x}/L$ and $\psi \equiv N_0 l_{\rm x}/L$.  
Again, since $I_{\rm PL} \propto N$ and $I_{\rm pump} \propto N_0$, $I_{\rm PL} = c_1 \zeta$ and $I_{\rm pump} = c_2 \psi$, where $c_1$ and $c_2$ are constants.  With $\zeta$ and $\psi$, Eq.~(\ref{dN0_dN}) can be written in a dimensionless form:
\begin{equation}
\frac{d \psi}{d \zeta} =
\frac{1}
{
(1 - \zeta)
\left\{
1 - \frac{l_{\rm x}}{L} {\left(1 - \zeta\right)}^{-1}
\right\}^{\frac{L}{l_{\rm x}} \zeta}
} \, .
\label{psi-zeta2}
\end{equation}
Furthermore, similar to the steady-state case, an expansion is performed on the second factor in the denominator of Eq.~(\ref{psi-zeta2}) together with elimination of the higher-order terms of $l_{\rm x}/L$, resulting in a differential equation of only $\zeta$ and $\psi$:
\begin{equation}
\frac{d \psi}{d \zeta} = \frac{1}{1-\zeta} \, \mathrm{exp}\left(\frac{\zeta}{1-\zeta} \right) \;.
\label{instantaneous-eqn-before-integ_(lambda=1)}
\end{equation}
Finally, by integrating Eq.~(\ref{instantaneous-eqn-before-integ_(lambda=1)}) from 0 to $\zeta$, we obtain the solution for $\lambda = 1$ ($ex + ex \rightarrow ex$) as
\begin{equation}
\psi = \frac{1}{e} \, \left\{ \mathrm{Ei}\left(\frac{1}{1-\zeta} \right) - \mathrm{Ei}(1) \right\}\;,
\label{instantaneous-eqn_(lambda=1)}
\end{equation}
where
\begin{equation}
\mathrm{Ei}(x) = \int_{-\infty}^{x} \frac{e^y}{y} dy
\label{exponential_integral}
\end{equation}
is the exponential integral.

Similarly, the solution for $\lambda = 0$ ($ex + ex \rightarrow 0$) in the instantaneous limit is
\begin{equation}
\psi =
\int_0^{\zeta}
\frac{1}{
2 (1 - \zeta) \mathrm{exp}\left( -\frac{1}{1-\zeta} \right) - 1
}
d \zeta
\label{instantaneous-eqn_(lambda=0)}
\end{equation}
where the integral has to be solved numerically.  Those dimensionless equations~(\ref{instantaneous-eqn_(lambda=1)}) and (\ref{instantaneous-eqn_(lambda=0)}) become equivalent ($\zeta = \psi$) in the low density limit ($\zeta \rightarrow 0$).

\section{Comparisons and Discussion}
\label{Discussion}

\subsection{Comparison of model and experiment}

To compare with the experimental data shown in Fig.~\ref{saturation}, we use Eq.~(\ref{instantaneous-eqn_(lambda=1)}), which is for the instantaneous limit with $\lambda = 1$.  The choice of this equation is because of the short duration of the optical pulses used ($\sim$250~fs), the very fast ($\sim$40~fs\cite{Manzoni-PRL-2005}) and efficient\cite{Hertel-NanoLett-2008,Lebedkin-PRB-2008} internal decay of excitons created at $E_{22}$ levels to the lowest $E_{11}$ level, and the much longer spontaneous decay time from the $E_{11}$ level to G.~S. (10--100~ps).\cite{Ostojic-PRL-2004,Wang-PRL-2004,Reich-PRB-2005}  The solid curves shown in Fig.~\ref{saturation} are fitting curves using Eq.~(\ref{instantaneous-eqn_(lambda=1)}), indicating that the model agrees well with the experimentally observed PL saturation curves.

Table \ref{parameters} is a summary of the fitting analysis performed on the data in Fig.~\ref{saturation} using Eq.~(\ref{instantaneous-eqn_(lambda=1)}).  The first two columns show the optimum values of the scaling factors $c_1$ and ${c_2}^{-1}$.  These are thought to be proportional to the oscillator strength for the PL emission from $E_{11}$ levels and the oscillator strength for the optical absorption around $E_{22}$ levels, respectively, as expected from $I_{\rm PL} \propto c_1 N$ and $N_0 \propto {c_2}^{-1} I_{\rm pump}$.  
The right three columns ($\zeta$, $N$, and $R$) are values at $1.02 \times 10^{14}$ photons/cm$^2$ (see Fig.~\ref{saturation}).  The values of $N$ were obtained through the relationship $N = \zeta /l_{\rm x}$, where $l_{\rm x}$ was assumed to be 45 nm (one half of the average exciton excursion range defined in Ref.~\onlinecite{Cognet-Science-2007}).  The exciton density in the highly saturated regime ($\sim 1 \times  10^5$ cm$^{-1}$) is still more than one order magnitude smaller than the expected Mott density in SWNTs ($\sim 7 \times 10^6$ cm$^{-1}$, assuming an exciton size of $\sim$1.5~nm\cite{Perebeinos-PRL-2004}), as has been discussed in detail in Ref.~\onlinecite{murakami-under-review}.  $R$ in the right-most column denotes the ratio between the number of as-created $e$-$h$ pairs and the number of excitons that survived the EEA process until their spontaneous decay to G.~S., or $N_0 / N$.  The values of $R$ show that approximately 90\% of the initially created $e$-$h$ pairs/excitons decay non-radiatively through the EEA path when the $E_{22}$ levels are resonantly excited at a fluence of $1.02 \times 10^{14}$ photons/cm$^2$.

As expected, the values of $c_1$ are similar within the same chirality type, regardless of the excitation wavelength. However, for the particular case of (6,5) with 570~nm excitation, where the excitation wavelength exactly coincided with the $E_{22}$ resonance peak, the value of $c_1$ ($\sim 3 \times 10^5$) is appreciably smaller than those at the other excitation wavelengths ($\sim 4 \times 10^5$).  Such a smaller $c_1$ value implies that more excitons decayed to the G.~S. through non-radiative paths compared to the other cases.  This may be due to an emergence of stronger nonlinear processes that have not been taken into account in the assumptions, such as the breakdown of the $\lambda = 1$ assumption and/or the appearance of \emph{more-than-two}-body annihilation processes because of the very high initial $e$-$h$ pair density, achieved with the strong $E_{22}$ resonance in this particular case.

The values of ${c_2}^{-1}$ also show the expected tendency toward higher (lower) values when the excitation wavelength is closer to (further from) the $E_{22}$ resonance wavelength.  The slightly higher ${c_2}^{-1}$ value for the case of (7,5) with 570~nm excitation is again attributed to its vicinity to the $E_{22}$ phonon sideband at $\sim$585~nm.\cite{Miyauchi-PRB-2006}

\begin{table}
\begin{tabular}{c|c||c|c|c|c|c} \hline
      &   Exct. & $c_1$ & ${c_2}^{-1}$ & $\zeta$ & $N$ &  $R$     \\
\raisebox{1.5ex}[0pt]{Type} &  (nm)
                & ($\times 10^5$) & ($\times 10^{-14}$) &     &  ($\times 10^5$ cm$^{-1}$) &      \\
\hline\hline
      &   570   &  2.97  &   7.74    &  0.764  &   1.70   &  10.3    \\
\cline{2-7}
(6,5) &   615   &  3.89  &   1.54    &  0.566  &   1.26   &  2.78    \\
\cline{2-7}
      &   654   &  4.21  &   1.13    &  0.505  &   1.12   &  2.28    \\
\hline\hline
      &   570   &  1.14  &   3.18    &  0.678  &   1.51   &  4.77    \\
\cline{2-7}
(7,5) &   615   &  1.34  &   2.68    &  0.656  &   1.46   &  4.17    \\
\cline{2-7}
      &   654   &  1.22  &   9.16    &  0.776  &   1.72   &  12.0    \\
\hline\hline
      &   570   &  0.73  &   1.54    &  0.566  &   1.26   &  2.78    \\
\cline{2-7}
(8,3) &   615   &  0.72  &   2.61    &  0.652  &   1.45   &  4.07    \\
\cline{2-7}
      &   654   &  0.68  &   5.05    &  0.729  &   1.62   &  7.05    \\
\hline
\end{tabular}
\caption{Summary of the optimal parameters to used fit the experimental data, obtained from the analysis with Eq.~(\ref{instantaneous-eqn_(lambda=1)}).
$\zeta$, $N$, and $R$ are values at the fluence of $1.02 \times 10^{14}$ photons/cm$^2$, corresponding to the largest fluence in Fig.~\ref{saturation}.
}
\label{parameters}
\end{table}

\subsection{Comparison with Monte Carlo calculation}

\begin{figure}
\includegraphics[scale=0.6]{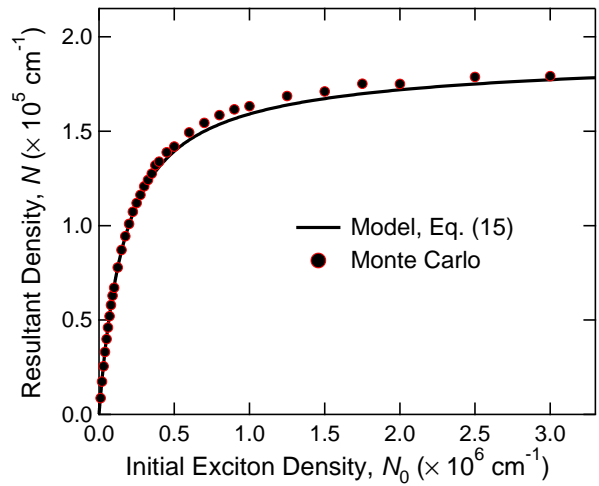}
\caption{(color online).
Comparison between the saturation behavior predicted by Eq.~(\ref{instantaneous-eqn_(lambda=1)}) (black solid curve) and the result obtained by a Monte Carlo simulation of the EEA process.  The abscissa denotes the density of initially created excitons while the ordinate corresponds to the resultant density. The experimental range of $N_0$ in Fig.~\ref{saturation} is $N_0 \leq 2 \times 10^6$ cm$^{-1}$. 
}
\label{simulation}
\end{figure}

To further confirm the validity of the model, we performed a computational simulation based on the Monte Carlo method.  At the beginning of the simulation, a random distribution of $N_0$ excitons along a line is created at time $t = 0$, corresponding to the instantaneous limit.  For $t > 0$, each exciton undergoes a random movement in each computational step $dt$ by the distance given by the probability distribution of $\bar{N}(0,l_0^2)$, where $\bar{N}(x_0, \sigma^2)$ denotes the normal distribution centered at $x_0$ with variance $\sigma$ and $l_0 \equiv \sqrt{D dt}$.  Upon intersection of any two excitons, the EEA of $\{ ex + ex \rightarrow ex \}$ takes place.  In addition, each exciton is eliminated from the system with a probability of $\gamma_{\rm tot} dt$ ($= {\tau_{\rm tot}}^{-1} dt$), corresponding to possible spontaneous decay during each computational step. The total density of excitons that have decayed spontaneously ($N$) can thus be calculated for each value of the initial density $N_0$.

Figure~\ref{simulation} shows a comparison between the model for the instantaneous limit [Eq.~(\ref{instantaneous-eqn_(lambda=1)})] with $l_{\rm x}$ = 45 nm and the result of the Monte Carlo simulation, plotting the relationship of $N_0$ and $N$. The simulation was performed with $D = 0.42$ cm$^2$/s and $\tau_{\rm tot} = 100$ ps,\cite{footnote-1} which resulted in an average exciton displacement of 45 nm based on the same simulation performed without EEA. The comparison shows satisfactory agreement, indicating that the simple analytical form of Eq.~(\ref{instantaneous-eqn_(lambda=1)}) well describes the phenomenon of diffusion-limited EEA, hence validating our model.


\subsection{Comparison with conventional rate-equation solution}

Conventionally, a rate equation of the form
\begin{equation}
\frac{d N(t)}{dt} = G_{\rm in}(t) - \gamma_{\rm tot} N(t) - \gamma_{\rm EEA} {N(t)}^2
\label{conventional_1}
\end{equation}
has been used to explain EEA processes observed for various materials from 1-D to 3-D.\cite{Valkunas-BPJ-1995,Roberti-JCP-1998,OHara-PRB-1999,Dicker-PRB-2005,Ma-PRL-2005,Zaushitsyn-PRB-2007,King-PRB-2007} The terms on the right hand side, from left to right, represent the rate of excitons entering the system (or the $E_{\rm 11}$ level in the present case), the rate of excitons spontaneously decaying from the system to the G.~S. (at $\gamma_{\rm tot}$), and the rate of excitons leaving the system because of EEA (at $\gamma_{\rm EEA}$), respectively.  
Here, the probability of finding two excitons at the same position (or the EEA rate) is assumed to be proportional to $N^2$, and hence, exciton diffusion as well as the finite size (i.e., length) that each 1-D exciton occupies are not taken into account in Eq.~(\ref{conventional_1}).

Equation~(\ref{conventional_1}) is solved for the pulse-wise creation of $N_0$ excitons at $t = 0$ [i.e., $G_{\rm in}(t) = \delta(N_0)$] in order to be compared with the proposed model for the instantaneous limit [Eq.~(\ref{instantaneous-eqn_(lambda=1)})] as well as with the experimental results. With this initial condition, Eq.~(\ref{conventional_1}) is readily solved to be
\begin{equation}
N(t) = \frac{1}
{
\left( \frac{1}{N_0} + \Gamma \right) \mathrm{exp}\left(\gamma_{\rm tot} t \right) - \Gamma
}
\; \; \left[ \Gamma \equiv \frac{\gamma_{\rm EEA}}{\gamma_{\rm tot}} \right] \;.
\label{conventional_2}
\end{equation}
The total number of photons emitted from the sample, $I_{\rm PL}$, is obtained by integrating Eq.~(\ref{conventional_2}) from $t = 0$ to $\infty$ as
\begin{equation}
I_{\rm PL}
= \eta \int_0^{\infty} \gamma_{\rm tot} N(t) dt
= - \frac{\eta}{\Gamma} \,\mathrm{ln}\left\{1 - { \left(1 + \frac{1}{N_0 \Gamma}\right) }^{-1} \right\}
\label{conventional_3}
\end{equation}
where $\eta = \gamma_{\rm r} / \gamma_{\rm tot}$ as before.  Equation~(\ref{conventional_3}) describes the relationship between $I_{\rm PL}$ and $N_0$ in terms of two adjustable components $\eta/\Gamma$ and $N_0 \Gamma$. 

The dashed lines drawn in Fig.~\ref{saturation} are the optimum fits by Eq.~(\ref{conventional_3}) to the experimental data. While Eq.~(\ref{conventional_3}) reproduces the behavior well for the regime of weaker PL saturation with non-resonant excitations, the deviation becomes clearer for the regime of stronger PL saturation with resonant excitations. Most importantly, however, the density predicted by Eq.~(\ref{conventional_3}) has \emph{no upper limit} as recognized by the steadily increasing dashed lines shown in Fig.~\ref{saturation} [or by the logarithmic form of Eq.~(\ref{conventional_3})], while Eq.~(\ref{instantaneous-eqn_(lambda=1)}) correctly demonstrates the existence of an upper limit, as observed experimentally and in the Monte Carlo simulation. 

Another important difference between Eq.~(\ref{instantaneous-eqn_(lambda=1)}) and Eq.~(\ref{conventional_3}) is described as follows:  The solution of the conventional rate equation [Eq.~(\ref{conventional_3})] contains two independent quantities: $\eta/\Gamma$ and $N_0 \Gamma$. The former can be used to scale $I_{\rm PL}$. However, since the parameter $N_0 \Gamma$ simultaneously scales $I_{\rm pump}$ ($\propto N_0$) \emph{and} changes the shape of the saturation curve, one essentially cannot estimate the density of excitons from the fitting analysis based on Eq.~(\ref{conventional_3}). On the other hand, since the shape of Eq.~(\ref{instantaneous-eqn_(lambda=1)}) has been uniquely determined by its dimensionless representation, the only thing one can do is to \emph{linearly and independently} change the two scaling constants $c_1$ and $c_2$ for $I_{\rm PL}$ and $I_{\rm pump}$, respectively, yielding sets of $\zeta$ and $\psi$ values along the scaled (or fitted) $I_{\rm PL}$ vs.~$I_{\rm pump}$ curves.


\section{Summary}
\label{Summary}

We have investigated photoemission properties of high-density 1-D excitons in single-walled carbon nanotubes.  As the excitation intensity increases, all photoluminescence emission peaks arising from different chiralities showed clear saturation in intensity.  Each peak exhibited a saturation value that was independent of the excitation wavelength, indicating that there is an upper limit on the exciton density for each nanotube species.  We interpret these results in terms of diffusion-limited exciton-exciton annihilation processes through which high-density excitons decay non-radiatively.  

To quantitatively understand the saturation behavior observed in the experiment, we have developed a theoretical model, taking into account the generation, diffusion-limited exciton-exciton annihilation, and spontaneous decays of 1-D excitons.  The saturation curve predicted by the model under appropriate approximations reproduced the experimental saturation curves well, and the fitting analysis allowed us to estimate the density of excitons for a given diffusion constant.  We also compared our results with Monte Carlo calculations, confirming the validity of our model. Additionally, we examined the saturation behavior predicted by the solution of the conventional exciton-exciton annihilation rate equation that does not take into account diffusion and showed that the solution qualitatively failed to fit the experimental data at high exciton densities, showing the inappropriateness of its use for excitons in carbon nanotubes.  The approach presented in this paper should have wide applicability for predicting the intensities of photoluminescence from single-walled carbon nanotubes under various excitation conditions.

An important conclusion drawn from the current study is that a large density of electron-hole pairs is difficult to achieve in single-walled carbon nanotubes.  This imposes a serious challenge for making active optoelectronic devices based on semiconducting single-walled carbon nanotubes --- 1-D semiconductor lasers, in particular, which would require densities comparable to, or higher than, the Mott density.  The existence of an upper limit on the density of excitons would also prevent fundamental studies of bosonic characters of 1-D excitons expected at quantum-degenerate densities.  Hence, novel methods are needed for minimizing non-radiative decay processes such as exciton-exciton annihilation in single-walled carbon nanotubes, which would require ways to control the dynamic parameters of excitons such as the spontaneous decay rate and diffusion constant through, e.g., varying the temperature as well as applying an external magnetic field.~\cite{Ajit-Condmat}
   

\begin{acknowledgments}
The authors thank Ajit Srivastava for valuable discussions and Erik Einarsson for emendation.
We thank the Robert A.~Welch Foundation (C-1509) and NSF (DMR-0325474 and OISE-0530220) for support.  One of us (Y.~M.) was financially supported by Grants-in-Aid for Scientific Research (\#18-09883) from the Japan Society for the Promotion of Science (JSPS). Y.~M.~thanks Tatsuya Okubo and Shigeo Maruyama for support to his fulfillment of the JSPS research program.
\end{acknowledgments}



\end{document}